\documentclass[aps,prd,reprint,superscriptaddress,showkeys,amsmath,amssymb,amsfonts]{revtex4-2}
\usepackage[T1]{fontenc}
\usepackage[utf8]{inputenc}
\usepackage[spanish,USenglish]{babel}
\usepackage[colorlinks,citecolor=blue,urlcolor=blue,linkcolor=blue]{hyperref}
\usepackage{braket}
\usepackage{mathtools}
\usepackage{enumerate}
\usepackage{dcolumn}
\usepackage{bm}

\makeatletter

\newenvironment{nobotrulewidetext}{%
  \par\ignorespaces
  \setbox\widetext@top\vbox{%
   \hb@xt@\hsize{%
    \leaders\hrule\hfil
    \vrule\@height6\p@
   }%
  }%
  \setbox\widetext@bot\hb@xt@\hsize{%
    \vrule\@depth6\p@
    \leaders\hrule\hfil
  }%
  \onecolumngrid
  \vskip10\p@
  \dimen@\ht\widetext@top\advance\dimen@\dp\widetext@top
  \cleaders\box\widetext@top\vskip\dimen@
  \vskip6\p@
  \prep@math@patch
}{%
  \par
  \twocolumngrid\global\@ignoretrue
  \@endpetrue
}%

\makeatother

\DeclarePairedDelimiter{\abs}{\lvert}{\rvert}
﻿
\begin{document}

\title{Diabatic description of charmoniumlike mesons II: mass corrections and strong decay widths}
\author{R. Bruschini}
\email{roberto.bruschini@ific.uv.es}
\affiliation{\foreignlanguage{spanish}{Unidad Teórica, Instituto de Física Corpuscular (Universidad de Valencia--CSIC), E-46980 Paterna (Valencia)}, Spain}
\author{P. González}
\email{pedro.gonzalez@uv.es}
\affiliation{\foreignlanguage{spanish}{Unidad Teórica, Instituto de Física Corpuscular (Universidad de Valencia--CSIC), E-46980 Paterna (Valencia)}, Spain}
\affiliation{\foreignlanguage{spanish}{Departamento de Física Teórica, Universidad de Valencia, E-46100 Burjassot (Valencia)}, Spain}


\begin{abstract}
From a diabatic bound state approach to $J^{PC}=1^{--}$ and $(0,1,2)^{++}$
charmoniumlike resonances below $4.1$ GeV, formulated in terms of
${c\overline{c}}$  and closed meson-meson channels,
we calculate mass shifts and widths due to open meson-meson channels.
This calculation does not involve any new free parameter, so comparison
of our predictions with existing data provides a direct test of our approach.
Further  mass corrections
are also estimated and good agreement with the measured masses comes out.
As for the calculated widths, overall reasonable, they point out to the need of some refinement
of our current bound state approximation for an accurate description of data.
These results give additional support to the diabatic approach in QCD as an adequate framework for a complete
unified description of conventional and unconventional charmoniumlike resonances. In this respect,
the experimental discovery of a predicted $2^{++}$ resonance  with a mass around  $4$ GeV
would be of special relevance.
\end{abstract}

\maketitle

\section{Introduction}

A current challenge in hadron physics is to achieve a QCD based description of
(unconventional) quarkoniumlike mesons such as $\chi_{c1}(3872)
$, $X(3915)$, and others \cite{PDG20} discovered in the last
two decades, whose properties do not correspond to a very dominant
conventional heavy quark $(Q)$ - heavy antiquark $(\overline{Q})$ meson structure.

Although there are nowadays compelling experimental indications of the
presence of additional open-flavor meson-meson components in some of these
uncoventional states, the QCD theoretical implementation of such components,
together with the $Q\overline{Q}$ ones for a complete quarkoniumlike
description, presents some difficulties (for a comprehensive review of
the experimental and theoretical situation, see
\cite{Bra11,Bod13,Chen16,Hosaka16,Chen17,Dong17,Esp17,Leb17,Guo18,Olsen18,Liu19,Yuan19, Bra20} and references
therein). Just recently, noticeable progress in this direction, based on
lattice QCD results, has been reported. More concretely, unquenched lattice
QCD calculations of the energies for static $Q$ and $\overline{Q}$ sources,
when the $Q\overline{Q}$ configuration mixes with one or two open-flavor
meson-meson configurations, have been carried out \cite{Bal05,Bul19}. These
static energies can be directly related to the potential matrix entering in a
multichannel Schr\"{o}dinger equation for the $Q\overline{Q}$ and meson-meson
components of a quarkoniumlike meson. This relation has been used to explore
the bottomoniumlike spectrum when only one (average) meson-meson channel is
taken into account \cite{Bic20}. A more general and systematic formulation of
this relation for any number of meson-meson channels has been recently
proposed \cite{Bru20}. This formulation translates the diabatic approach used
in molecular physics to tackle the electronic configuration mixing problem
\cite{Bae06} to the study of quarkoniumlike systems.

The direct diabatic correspondence between lattice QCD calculations
and the potential matrix entering the Schr\"{o}dinger equation
constitutes a substantial improvement over existing descriptions
of quarkoniumlike mesons where the interaction between $Q\overline{Q}$
and meson-meson degrees of freedom has no clear connection with QCD.

It is worth to emphasize that the diabatic approach goes beyond the
(single-channel) Born-Oppenheimer approximation in QCD developed in
\cite{Jug99} and applied in \cite{Braa14} to the description of quarkonium and
quarkonium hybrids. Indeed, the diabatic approach reduces to
the Born-Oppenheimer approximation for conventional quarkonium $Q\overline{Q}$
for energies far below the lowest open-flavor meson-meson threshold,
but unlike the Born-Oppenheimer approximation it maintains its validity for energies
close below or above that threshold. We refer to the previous paper of this series \cite{Bru20} for
a more detailed explanation about the advantages of
the diabatic approach over the Born-Oppenheimer approximation.

Therefore, the so-called diabatic approach in QCD provides an appropriate framework
for a unified and complete nonperturbative description of conventional
quarkonium and quarkoniumlike mesons made of $Q\overline{Q}$ and meson-meson components.

In the first practical application of the diabatic formalism to the
description of charmoniumlike mesons we have considered them stable against
decays into open-flavor meson-meson channels, neglecting the fact that states
above any possible meson-meson threshold with the same quantum numbers are
unstable. This approximation has been implemented by solving the multichannel Schr\"{o}dinger equation taking
into account that a bound state solution with a correct asymptotic behavior
requires that meson-meson components with thresholds below the mass of the
bound state be neglected. This description, that has been applied to isoscalar $I=0$ states with
$J^{PC}=1^{--}$ and $(0,1,2)^{++}$ below $4.1$ GeV, is expected
to be well suited for narrow states as the ones below the first absolute
meson-meson threshold $D\overline{D}$, or the first excited $1^{++}$ state,
which is assigned to $\chi_{c1}(3872)$, for which there is no
any coupled lower threshold. For the other states the treatment may also be
suited if mass corrections due to the neglected meson-meson components and
widths can be properly incorporated.

In this article we go a step further in the diabatic description of
quarkoniumlike mesons by evaluating, from the calculated bound state
solutions, the contribution of the neglected meson-meson components to their
masses and decay widths. Then, a more detailed comparison with data may be
done. In this regard, one should always keep in mind that there might still
exist experimental quarkoniumlike resonances escaping from the
bound state approximation we follow.

The contents of the article are organized as follows. In Sec.~\ref{sec2} a brief
resume of the diabatic bound state description of quarkoniumlike mesons is
presented. In Sec.~\ref{sec3} we detail the framework for the calculation of mass
corrections and widths to open-flavor meson-meson channels. The results
obtained for charmoniumlike systems are listed and discussed in Sec.~\ref{sec4}.
Finally, in Sec.~\ref{sec5} we summarize our main results and conclusions.

\section{\label{sec2}Diabatic bound state description of quarkoniumlike mesons}

The diabatic approach in QCD has been detailed in \cite{Bru20}. Quarkoniumlike
mesons, characterized by quantum numbers $J^{PC}$, are assigned to solutions of the multichannel Schr\"{o}dinger equation
\begin{equation}
(\mathrm{K}+\mathrm{V}(r))\Psi(
\bm{r})=E\Psi(\bm{r})\label{CM}
\end{equation}
where $\Psi(\bm{r})$ is a column vector notation for the wave function,
\begin{equation}
\Psi(\bm{r})=
\begin{pmatrix}
\psi_{Q\overline{Q}}(\bm{r}) \\
\psi_{M\overline{M}}^{(1)}(\bm{r}) \\
\vdots\\
\psi_{M\overline{M}}^{(N)}(\bm{r})\\
\end{pmatrix},
\end{equation}
with $\psi_{M\overline{M}}^{(i)}(\bm
{r})$, $i=1,2\dots$ standing for the $i$-th meson-meson component.

$\mathrm{K}$ is the kinetic energy matrix
\begin{equation}
\mathrm{K}=
\begin{pmatrix}
-\frac{1}{2\mu_{Q\overline{Q}}}\nabla^{2} &  &  & \\
& -\frac{1}{2\mu_{M\overline{M}}^{(1)}}\nabla^{2} &  & \\
&  & \ddots & \\
&  &  & -\frac{1}{2\mu_{M\overline{M}}^{(N)}}\nabla^{2}
\end{pmatrix}
\label{Kinetic}
\end{equation}
where $\mu_{M\overline{M}}^{(i)}$ is the reduced mass of the $i$-th
meson-meson component, and matrix elements equal to zero are not displayed.

$\mathrm{V}(r)$ is the diabatic potential matrix
\begin{equation}
\mathrm{V}(r)=
\begin{pmatrix}
V_{\text{C}}(r) & V_{\textup{mix}}^{(1)}(r) & \hdots & V_{\textup{mix}}^{(N)}(r)\\
V_{\textup{mix}}^{(1)}(r) & T_{M\overline{M}}^{(1)} &  & \\
\vdots &  & \ddots & \\
V_{\textup{mix}}^{(N)}(r) &  &  & T_{M\overline{M}}^{(N)}
\end{pmatrix}.
\label{dpotmany}
\end{equation}
The diagonal term $V_{\text{C}}(r)$ stands for the $Q\overline{Q}$ Cornell
potential
\begin{equation}
V_{\text{C}}(r)=\sigma r-\frac{\chi}{r}+ m_{Q}+ m_{\overline{Q}
}-\beta\label{CPOT}
\end{equation}
with $\sigma$, $\chi$, $m_{Q}$ and $\beta$ being the string tension, the
colour coulomb strength, the heavy quark mass, and a constant fixing the
origin of the potential respectively.

Any of the other diagonal terms stands for the mass of a meson-meson
threshold
\begin{equation}
T_{M\overline{M}}^{(i)}= m_{M}^{(i)}+ m_{\overline{M}}^{(i)}
\end{equation}
with $m_{M}^{(i)}$ and $m_{\overline{M}}^{(i)}$ being the masses of the
corresponding meson and antimeson.

The off-diagonal term $V_{\textup{mix}}^{(i)}(r)$ stands for the mixing
potential between the $Q\overline{Q}$ and the $i$-th meson-meson component,
that can be calculated \textit{ab initio} from lattice QCD.

Notice that no interaction potential between different meson-meson components is
considered, what it is justified for isolated, well separated meson-meson
thresholds with no overlap at all (including the meson widths). Notice though
that an indirect interaction through their coupling to the $Q\overline{Q}$
channel is present.

Moreover, for the spherically-symmetric and spin-independent diabatic potential
matrix we use, each $Q\overline{Q}$ configuration with a distinct value of
$(l_{Q\overline{Q}},s_{Q\overline{Q}})$ is taken as a channel \textit{per se},
and the same for each meson-meson configuration with a distinct value of
$(l_{M\overline{M}}^{(i)},s_{M\overline{M}}^{(i)
})$. The possible values of $(l,s)$ are determined by the $J^{PC}$ quantum
numbers of the quarkoniumlike system under consideration. Then, for a given
$J^{PC}$ the mixing potential entering the coupled radial equations,
$V_{\textup{mix}}^{(i)}(r)$, is the same between any one of the $(l_{Q\overline
{Q}},s_{Q\overline{Q}})$ and any one of the $(l_{M\overline{M}}^{(
i)},s_{M\overline{M}}^{(i)})$ channels.


In order to calculate bound states some constraints are imposed.
First, a finite number of thresholds is considered. This is justified because
for a given bound state the probability of meson-meson components
corresponding to thresholds far above the mass of the bound state is negligible.
Additionally, only meson-meson components above the bound state mass are considered.
Otherwise the free-wave behavior introduced by meson-meson components with
threshold below the mass of the bound state would prevent us from obtaining a
physical (bound state) solution. In practice this is done by selecting a
subset of thresholds, from a given threshold up, and calculating the bound
states below it. In order to avoid a possible double counting when different
threshold selections are considered a one to one correspondence between the
Cornell $Q\overline{Q}$ and the quarkoniumlike bound states is assumed.

Notice that a solution of the Schr\"{o}dinger equation above threshold is
possible, but the eigenstates would represent meson-meson scattering states. In such a
context quarkoniumlike mesons would be described as resonances in a properly
defined meson-meson scattering problem. A development of a diabatic
description of the coupled channel meson-meson scattering, out of the scope of
the present article, is underway.


The diabatic bound state approach has been applied to isoscalar $I=0$
charmoniumlike $(Q=c)$ mesons with $J^{PC}=1^{--}$ and $(
0,1,2)^{++}$ below $4.1$ GeV for which the threshold-threshold
interactions can be safely neglected. For the Cornell potential \eqref{CPOT},
standard values of the parameters
\begin{subequations}
\label{params}
\begin{align}
\sigma &  =925.6\text{~MeV/fm},\\
\chi &  =102.6\text{~MeV~fm},\\
m_{c} &  =1840\text{~MeV}\\
\beta &  =855\text{~MeV}
\end{align}
\end{subequations}
have been chosen \cite{Eic94}.

The meson-meson thresholds, calculated from the masses of charmed and charmed strange
mesons in \cite{PDG20}, are listed in Table~\ref{tablist}. It is worth to remark that
the use of the masses of the experimental thresholds introduces some implicit
spin dependence in the description.


\begin{table}
\begin{ruledtabular}
\begin{tabular}{cc}
$M\overline{M}$		& $T_{M\overline{M}}$ (MeV)		\\
\hline
$D\overline{D}$ 			& $3730$								\\
$D\overline{D}^\ast(2007)$ 		& $3872$ 							\\
$D_{s}^{+}D_{s}^{-}$			& $3937$							\\
$D^\ast(2007)\overline{D}^\ast(2007)$	& $4014$						\\
$D_{s}^{+}D_{s}^{\ast-}$ 		& $4080$							\\
\end{tabular}
\end{ruledtabular}
\caption{\label{tablist}Low-lying open-charm meson-meson thresholds.
Threshold masses $T_{M\overline{M}}^{(i)}$ from the measured charmed and charmed
strange meson masses \cite{PDG20}.}
\end{table}


The possible values of $l_{c\overline{c}}$ and $l_{M\overline{M}}^{(
i)}$ contributing to a given set of quantum numbers $J^{PC}$ are
shown in Table~\ref{quanum} where the common notation $D_{(s)}$ to refer to charmed
as well to charmed strange mesons is used.


\begin{table}
\begin{ruledtabular}
\begin{tabular}{ccccc}
$J^{PC}$	& $l_{c\overline{c}}$	& $l_{D_{(s)} \overline{D}_{(s)}}$	& $l_{D_{(s)} \overline{D}_{(s)}^\ast}$	& $l_{D_{(s)}^{\ast} \overline{D}_{(s)}^\ast}$ 	\\
\hline
$0^{++}$	& $1$		& $0$						& 								& 0, 2						\\
$1^{++}$	& $1$		& 							& $0, 2$							& $2$						\\
$2^{++}$	& $1, 3$		& $2$						& $2$							& $0, 2, 4$					\\
$1^{--}$	& $0, 2$		& $1$						& $1$							& $1, 3$						\\
\end{tabular}
\end{ruledtabular}
\caption{\label{quanum}Values of $l_{c\overline{c}}$ and $l_{M\overline{M}}^{(i)}$ corresponding to definite values of $J^{PC}$. A missing entry for $l_{M\overline{M}}^{(i)}$ means that the particular meson-meson configuration cannot form a state with the corresponding quantum numbers.}
\end{table}


As for the mixing potential, the current dearth of unquenched lattice data for
charmoniumlike systems \cite{Bal11,Bal18} prevents its \textit{ab initio} calculation. Instead, its form has
been parametrized from lattice results for bottomoniumlike systems
\cite{Bal05,Bul19}. The reasons for this parametrization have been detailed in
\cite{Bru20}. It reads
\begin{equation}
\abs{{V_{\textup{mix}}^{(i)}(r)}}  = \frac{\Delta}{2}
\exp \biggl\{-\frac{(V_{\text{C}}(r)-T_{M\overline{M}}^{(i)})
^{2}}{2\sigma^{2}\rho^{2}}\biggr\}
\end{equation}
or equivalently, at distances for which $V_\textup{C}(r) \approx \sigma r + m_{Q} + m_{\overline{Q}} - \beta$,
\begin{equation}
\abs{{V_{\textup{mix}}^{(i)}(r)}} \approx \frac{\Delta}{2}\exp\biggl\{  -\frac{(r-r_\textup{c}^{(
i)})^{2}}{2\rho^{2}}\biggr\}
\end{equation}
where $\frac{\Delta}{2}$ stands for the strength and $\rho$, the width of the
second gaussian, for the radial scale of the mixing, whereas $r_\textup{c}^{(
i)}$ denotes the crossing radius defined by
\begin{equation}
V_{\textup{C}}(r_{\textup{c}}^{(i)})=T_{M\overline{M}}^{(i)} .
\end{equation}
Let us realize that the mixing is only effective in an interval around
$r_\textup{c}^{(i)}$ determined by the value of $\rho$. The numerical
values of $\rho$ and $\Delta$:
\begin{subequations}
\begin{align}
\Delta &  =130~\text{MeV},\\
\rho &  =0.3~\text{fm}
\end{align}
\end{subequations}
were obtained in \cite{Bru20} by fitting the mass of $\chi_{c1}(3872)$ via a
calculation involving only the $D\overline{D}^{\ast}$ threshold. In the
present calculation we have also included the tiny effect of the $D^{\ast
}\overline{D}^{\ast}$ threshold, therefore the use of the same values of the
parameters gives rise to a very little discrepancy between the calculated mass
of $\chi_{c1}(3872)$ and its measured value.


The calculated spectrum of $(0,1,2)^{++}$ states, containing
one $c\overline{c}$ component with $l_{c\overline{c}}=1$ ($1p$) or two
$c\overline{c}$ components with $l_{c\overline{c}}=1,3$ ($2p$ and $1f)$ and
$1^{--}$ states, containing one $c\overline{c}$ component with $l_{c\overline
{c}}=0$ ($1s)$ or two $c\overline{c}$ components with $l_{c\overline{c}}=0,2$
($2s$ and $1d$) is shown in Table~\ref{charm_joint_table}.


\begin{table*}
\begin{ruledtabular}
\begin{tabular}{ccddddddd}
$J^{PC}$	& Mass (MeV)	& c\overline{c}	& D \overline{D}	& D \overline{D}^\ast
	& D_s \overline{D}_s	& D^\ast \overline{D}^\ast	& D_s \overline{D}_s^\ast	&  D_s^\ast \overline{D}_s^\ast	\\
\hline
$0^{++}$	& 3508.8			& 99 \%		& 1 \%	& 		& 		& 		& 		&		\\
		& 3918.9			& 60\%		& 		& 		& 35 \%	& 5 \%	& 		&		\\
$1^{++}$	& 3509.8			& 100 \%		& 		& 		& 		& 		&		& 		\\
		& 3871.5			& 5 \%		& 		&  95 \%	& 		& 		&		& 		\\
$2^{++}$	& 3508.7			& (100, 0) \%	& 		& 		& 		& 		& 		&		\\
		& 3909.0			& (69, 8) \%	& 		& 		& 7 \%	& 13 \%	& 1 \%	& 2 \%	\\
		& 4006.6			& (18, 53) \%	& 		& 		& 		& 28 \%	& 1 \%	&		\\
$1^{--}$	& 3082.4			& (100, 0) \%	& 		& 		& 		& 		& 		&		\\
		& 3658.8			& (92, 1) \%	& 4 \%	& 1 \%	& 1 \%	& 1 \%	& 		&		\\
		& 3785.8			& (1, 95) \%	& 		& 2 \%	& 1 \%	& 1 \%	& 		&		\\
\end{tabular}
\end{ruledtabular}
\caption{\label{charm_joint_table}Masses, $c\overline{c}$ probabilities, and meson-meson probabilities, for $(0,1,2)^{++}$ and $1^{--}$
charmoniumlike states calculated neglecting interaction with open thresholds. The $c\overline{c}$ probabilities
from different values of $l_{c\overline{c}}$ (see Table~\ref{quanum}) have been reported
inside parentheses. A missing entry under a meson-meson configuration
means that the corresponding component gives negligible (i.e., probability inferior to $1\%$) or no contribution at all to the state.}
\end{table*}


Let us note that the content of Table~\ref{charm_joint_table} differs from the one of Tables~V
and VI in \cite{Bru20} due to some technical improvement in the numerical calculation, to the inclusion
of $l_{c\overline{c}}=3$ and $l_{M\overline{M}}^{(i)}=4$
channels what allows for the prediction of a new $2^{++}$ resonance (see
below), and to the exclusion of the $1^{--}$ excited state
assigned to $\psi(4040)$. The reason for this exclusion is that
this state is peculiar in the sense that the improved calculation puts the
state below the $D^{\ast}\overline{D}^{\ast}$ threshold preventing the decay
to any $D^{\ast}\overline{D}^{\ast}$ channel. However, the mass correction, calculated as
explained in the next section, makes the mass to be so close to the $D^{\ast
}\overline{D}^{\ast}$ threshold that a small variation in the parameters (well
within our expected uncertainty) puts the state above threshold permitting
such decay as experimentally observed. This instability suggests that a refinement of our approximation is
necessary for an appropriate description of $\psi(4040)$. This refinement may
require a more complete treatment incorporating all meson-meson channels on
the same foot (as in a diabatic coupled meson-meson scattering framework not
yet developed), and interactions between the bound states through their
couplings to the same thresholds. It may also have to do with an improvement
of the diabatic potential, through the inclusion of spin dependent terms and
the consideration of a differentiated mixing for hidden strange thresholds.

For a meaningful comparison with data we have to take into account that our
Cornell potential does not contain spin-dependent terms. Then, for pure
$(nl)$ $c\overline{c}$ states the calculated masses have to be
compared to the $(nl)$ experimental centroids; in the other
cases, where meson-meson components are present, since they are specific for a
set of $J^{PC}$ quantum numbers, the comparison has to be done with the
experimental candidates with the same $J^{PC}$.

As can be checked the calculated masses for the $(0,1,2)^{++}$
ground states, which are practically pure
charmonium states, are pretty close to the experimental ground state centroid
from $1P_{J}$ states at $3525.30\pm0.11$~MeV. As for the calculated masses of
the first excited $(0,1,2)^{++}$ states, all containing
significant meson-meson components, can be put in good correspondence with the
existing experimental candidates. So, $\chi_{c1}(3872)$, with a
measured mass of $3871.69\pm0.17$~MeV has been assumed to correspond to the
first excited $1^{++}$ state, what implies that it is essentially $(
95\%)$ a $D\overline{D}^{\ast}$ bound state; $\chi_{c_{0}}(3860)$,
with a measured mass of $3862_{-32-13}^{+26+40}$~MeV, corresponds to the first
excited $0^{++}$ state with an important $(35\%)$
$D_{s}\overline{D}_{s}$ component; and $\chi_{c2}(3930)$ with a
measured mass of $3922.2\pm1.0$~MeV corresponds to the first excited $2^{++}$
state with small but non negligible $(7\%)$ $D_{s}\overline
{D}_{s}$ and $(13\%)$ $D^{\ast}\overline{D}^{\ast}$ components.
It should also be remarked that the calculated masses for the first excited
$(0,2)^{++}$ states are in complete agreement with data regarding their
positions with respect to the $D_{s}\overline{D}_{s}$ threshold, both below it.

There is an additional $J^{++}$ experimental resonance below $4.1$ GeV, the
$X(3915)$, with a measured mass of $3918.4\pm1.9$~MeV, whose quantum numbers
are not well established being possible a $0^{++}$ as well as a $2^{++}$
assignment. In \cite{Bru20} we tentatively identified it with $\chi_{c_{0}
}(3860)$ but observed through the ($J/\psi\,\omega$)
decay channel. However, a $2^{++}$ identification with $\chi_{c2}(
3930)$ is also feasible. In this regard, the quantitative analysis of
the decay widths to open-flavor meson-meson channels may sheed some light on
its correct assignment.

There is also an additional theoretically predicted resonance corresponding to
the second $2^{++}$ excited state with no existing experimental candidate to
be compared with. This prediction is quite robust in the sense that it comes
out from the presence of the $1f$ Cornell state at $4034$~MeV.
Hence, the experimental finding
of this resonance would provide a strong support to our treatment. In this
respect, the analysis of its dominant strong decay widths that we carry out
later on may guide experimental searches.

As for $1^{--}$ a good correspondence of the calculated masses with data can
be established as well. So, the ground state is a conventional $(
100\%)$ $c\overline{c}$ charmonium $1s$ state, its calculated mass
being pretty close to the experimental $1s$ centroid at $3068.65\pm0.13$~MeV
(from $J/\psi$ and $\eta_{c}(1s)$).

For the first excited state, containing a very big $(92\%)$
$2s$ $c\overline{c}$ component and a quite small $(4\%)$
$D\overline{D}$ component, the calculated mass is close to the measured mass
of $\psi(2s)$: $3686.097\pm0.010$~MeV. The second excited state
is very dominantly a $1d$ $c\overline{c}$ state $(95\%)$, its
calculated mass comparing well with the mass of $\psi(3770)$:
$3778.1\pm0.7$~MeV.


It is worth to point out that we expect the calculated wave functions and
masses for states with a non-vanishing $D_{s}\overline{D}_{s}$ and
$D_{s}\overline{D}_{s}^{\ast}$ probability to be more uncertain that the ones
calculated for states with no such components. The reason is that the
parameters of the mixing potential have been fixed from the $D\overline
{D}^{\ast}$ threshold (through the fitting of the mass of $\chi_{c1}(
3872)$). Then, although the use of the same parameters for the
$D\overline{D}$ and $D^{\ast}\overline{D}^{\ast}$ thresholds, differing only
from $D\overline{D}^{\ast}$ in their spins, is justified in our spin
independent treatment, it may not be the case for $D_{s}\overline{D}_{s}$ and
$D_{s}\overline{D}_{s}^{\ast}$. For these thresholds the interaction with
$c\overline{c}$ involves the creation of a strange quark-antiquark pair so
that the strength and radial scale for the mixing may differ from the one
associated to the light quark $(u,d)$ pair creation involved in
the $D\overline{D}^{\ast}$\nobreakdash{-}$c\overline{c}$ interaction. Unfortunately, the lack
of
 detailed data clearly indicating the presence of
$D_{s}\overline{D}_{s}$ and $D_{s}\overline{D}_{s}^{\ast}$ components in some
experimental resonance prevents us from fixing the specific mixing parameters
for these thresholds and from estimating quantitatively the uncertainty
deriving from using the same parameters as for the other thresholds.
Notwithstanding this some qualitative prescription can be formulated in the
sense that the bound state masses and wave functions are better described if
not involving $D_{s}\overline{D}_{s}$ and $D_{s}\overline{D}_{s}^{\ast}$
components or involving a very small percentage of them.


Keeping this in mind, the reasonable agreement of the calculated masses with existing data for all
the excited states suggests that the diabatic bound state approximation to
charmoniumlike mesons, below and above threshold, may be appropriate and
constitutes a good starting point for the evaluation of mass corrections and
widths. Next we center on the mass shifts and widths having to do with the
neglected thresholds. As we shall prove they cannot be calculated using
perturbation theory against our previous assumption in \cite{Bru20}.

\section{\label{sec3}Mass corrections and widths}

In order to evaluate the effect of the $n$ neglected thresholds in the
calculation of bound states with masses above these thresholds, let us start
from the Hamiltonian
\begin{equation}
H_{0}= K+ V_{0}
\end{equation}
whose matrix representation in configuration space is given by the sum of $\mathrm{K}$  from \eqref{Kinetic} and
\begin{equation}
\mathrm{V}_{0}(r)=\left(
\begin{smallmatrix}
V_{\text{C}}(r) &  &  &  & V_{\textup{mix}}^{(n+1)}(r) & \hdots & V_{\textup{mix}
}^{(N)}(r)\\
& T_{M\overline{M}}^{(1)} &  &  &  &  & \\
&  & \ddots &  &  &  & \\
&  &  & T_{M\overline{M}}^{(n)} &  &  & \\
V_{\textup{mix}}^{(n+1)}(r) &  &  &  & T_{M\overline{M}}^{(n+1)} &  & \\
\vdots &  &  &  &  & \ddots & \\
{V_{\textup{mix}}^{(N)}(r)} &  &  &  &  &  & T_{M\overline{M}}^{(N)}
\end{smallmatrix}\right) .
\end{equation}
The wave functions, corresponding to solutions of the Schr\"{o}dinger equation
\begin{equation}
H_{0}\ket{\Psi_{0}}=E_{0}\ket{\Psi
_{0}}\label{USE}
\end{equation}
can be expressed as
\begin{equation}
\Psi_{0}(\bm{r})=
\begin{pmatrix}
\psi_{Q\overline{Q}}(\bm{r}) \\
\widetilde{\psi}_{M\overline{M}}^{(1)}(\bm
{r}) \\
\vdots\\
\widetilde{\psi}_{M\overline{M}}^{(n)}(\bm
{r}) \\
\psi_{M\overline{M}}^{(n+1)}(\bm{r})
\\
\vdots\\
\psi_{M\overline{M}}^{(N)}(\bm{r}) \\
\end{pmatrix}
\end{equation}
where we have used the notation $\widetilde{\psi}_{M\overline{M}}$ to
indicate the non coupled (free) meson-meson components.

Let us realize that any bound state obtained when neglecting the
thresholds with $j=1,\dots,n$ corresponds to a solution of \eqref{USE} of the type
\begin{equation}
\Psi_{\textup{bound}\,(N-n)}^{(k)}(
\bm{r})=
\begin{pmatrix}
\psi^{(k)}_{Q\overline{Q}}(\bm{r}) \\
\\
\\
\\
\psi_{M\overline{M}}^{(n+1)(k)}(\bm{r})\\
\vdots\\
\psi_{M\overline{M}}^{(N)(k)}(\bm{r}) \\
\end{pmatrix}
\end{equation}
with eigenvalue $M^{(k)}_{(N-n)}$ given by the calculated bound state mass, where $k$ stands
for a set of quantum numbers labeling the bound state.

On the other hand the free meson-meson states for any of the
meson-meson pairs in the channels $1,\dots,n$ correspond to solutions of \eqref{USE} of the type
\begin{equation}
\Psi_{\textup{free}(j)}^{(\bm{p}, s)}(\bm{r})
=
\begin{pmatrix}
\\
\\
\widetilde{\psi}_{M\overline{M}}^{(j)(\bm{p}, s)}(\bm{r}) \\
\\
\\
\\
\\
\end{pmatrix}
\end{equation}
with eigenvalues $
E_{(j)}^{(\bm{p})}= T_{M \overline{M}}^{(j)} +\frac{\bm{p}^{2}}{2 \mu_{M \overline{M}}^{(j)}}$, being $\bm
{p}$ the relative meson-meson momentum and $s$ the total meson-meson spin.
Notice that there is a continuum of eigenstates for each value of $j$.

The interaction Hamiltonian between bound
and free states is provided by those mixing terms in the diabatic potential matrix $V$ that have been discarded
in the construction of $V_{0}$:
\begin{equation}
H_{I} \equiv V - V_{0}.
\end{equation}
Its matrix representation in configuration space reads
\begin{equation}
\mathrm{H}_{I}(r)=
\begin{pmatrix}
& V_{\textup{mix}}^{(1)}(r) & \hdots & {V_{\textup{mix}}^{(n)}(r)} & & & \\
V_{\textup{mix}}^{(1)}(r) &  &  &  &  &  & \\
\vdots &  &  &  &  &  & \\
{V_{\textup{mix}}^{(n)}(r)} &  &  &  &  &  & \\
\\
\\
\end{pmatrix} ,
\label{hamix}
\end{equation}
where it can be seen that $H_{I}$ couples he $Q\overline{Q}$ component $\psi_{Q\overline{Q}}(
\bm{r})$ of the bound state and the free meson-meson channels
$\widetilde{\psi}_{M\overline{M}}^{(j)}(\bm{r})$ with $j=1,\dots,n$.

In order to study the effect of $H_{I}$ we have to realize that the
exact coincidence of the energy (mass) of the bound state with one energy of
the continuum for any of the $M^{(j)}\overline{M}^{(j)}$ components makes the
ordinary perturbation theory inadequate (see the appendix for a detailed
explanation). Instead, a procedure based on the solution of the
Schr\"{o}dinger equation for $H_{0} + H_{I}$ has been
developed in atomic and nuclear physics, see for instance \cite{Fan61}, and in
hadronic spectroscopy \cite{Eic80}.

In general, for a given set of quantum numbers $J^{PC}$, this procedure
involves the diagonalization of a matrix containing the interactions between
bound states induced through their coupling to the same meson-meson states.
However, as detailed in what follows the bound states we start from have been
calculated from Hamiltonians involving different sets of neglected thresholds.
Hence we can only do a consistent calculation if the meson-meson channel
induced interactions between different bound states are not taken into
account. Then, the problem gets reduced to the calculation of the corrections
to one bound state due to its interaction with the specific neglected thresholds.

To be more precise let us consider $Q=c$. For $J^{PC}=0^{++}$ there is one
bound (ground) state when no thresholds are neglected and two bound states
(ground and first excited) when the threshold $D\overline{D}$ is neglected.
The ground state is $100\%$ $c\overline{c}$ in both cases what means that it
is decoupled from any threshold. Hence, we restrict our study to one $0^{++}$
(first excited) bound state interacting with the $D\overline{D}$ threshold.
Similarly, for the first excited $2^{++}$ state obtained when neglecting the
$D\overline{D}$ and $D\overline{D}^{\ast}$ thresholds, we calculate the
corrections to the $2^{++}$ (first excited) bound state interacting with
$D\overline{D}$ and $D\overline{D}^{\ast}$. As for the second excited $2^{++}$
state, obtained when neglecting the $D\overline{D}$, $D\overline{D}^{\ast}$
and $D_{s}\overline{D}_{s}$ thresholds, we calculate its corrections due to
these thresholds forgetting about its interaction, induced through its
coupling to $D\overline{D}$, with the first excited bound state obtained with
the same Hamiltonian (notice that this first excited bound state differs from
the previous one calculated without neglecting $D_{s}\overline{D}_{s}$).

For the $1^{--}$ states we follow exactly the same protocol. So, for the second excited state we
calculate its corrections due to $D\overline{D}$.


Having reduced the calculation of the corrections to the one bound state case,
the physical effect of the coupling to the continuum is to dilute the bound
state through a band of stationary scattering states. The expressions for the
mass and width of the resulting resonance can be
found in \cite{Fan61} and \cite{Eic80}. Let us detail them for one bound state
and one threshold. Let $\ket{A}$ be the bound state with
quantum numbers $J^{PC}$, and $\ket{BC}$ a free
meson-meson state. In the center of mass frame the momentum of the resonance vanishes
and
\begin{subequations}
\begin{align}
E_{B} &= m_{B} + \frac{p_{BC}^{2}}{2 m_{B}} \\
E_{C} &= m_{C} + \frac{p_{BC}^{2}}{2 m_{C}}
\end{align}
\end{subequations}
with
\begin{equation}
p_{BC}^{2}=2\mu_{BC}(E_{BC}- m_{B} - m_{C})
\end{equation}
and $E_{BC}\equiv E_{B}+E_{C}$.

Then the mass correction is written as
\begin{multline}
\mathrm{M}-\mathrm{M}_{A}\\
=\sum_{\substack{m_{s_{B}},\\m_{s_{C}}}} \mathcal{P}\!\!\!\int
\mathrm{d}\bm{p}_{BC}\,\frac{\abs{\bra{
\bm{p}_{BC};s_{B},m_{s_{B}},s_{C},m_{s_{C}
}} H_{I}\ket{A}}
^{2}}{\mathrm{M}-E_{BC}}
\label{rawmcorr}
\end{multline}
where $\mathrm{M}$ is the mass of the resonance, $H_{I}$ stands for the interaction Hamiltonian operator, and $\mathcal{P}\!\!\int$ indicates the
Cauchy principal value integral.

To calculate the matrix element in the numerator we manipulate it as
\begin{multline}
\bra{\bm{p}_{BC};s_{B},m_{s_{B}}
,s_{C},m_{s_{C}}} H_{I}\ket{A}
\\=\sum_{s,m_{s}} C_{s_{B}, s_{C},s}^{m_{s_{B}}, m_{s_{C}}, m_{s}}
\bra{\bm{p}_{BC};s_{B},s_{C}
,s,m_{s}}\mathrm{H}_{I}\ket{A}.
\end{multline}
The wave function for the meson-meson state reads
\begin{equation}
\begin{split}
&\braket{\bm{p}_{BC};s_{B},s_{C}
,s,m_{s} \vert \bm{r}}\\
&=\frac
{e^{-i\bm{p}_{BC}\cdot\bm{r}}}{(
2\pi)^{\frac{3}{2}}}\xi_{s}^{m_{s}\dagger}\\
&= \sum_{l,m_{l}}\sqrt{\frac{2}{\pi}}i^{-l}j_{l}(p_{BC
}r)Y_{l}^{m_{l}}(\widehat{\bm{p}}_{BC})
Y_{l}^{m_{l}\ast}(\widehat{\bm{r}})\xi_{s}^{m_{s}\dagger}\\
&  =\sum_{l,m_{l}}\sqrt{\frac{2}{\pi}}i^{-l}j_{l}(p_{BC
}r)Y_{l}^{m_{l}}(\widehat{\bm{p}}_{BC}) \\
&\quad \times
\sum_{J^{\prime},m_{J}^{\prime}} C_{l, s, J^{\prime}}^{m_{l}, m_{s}, m_{J}^{\prime}}
\mathcal{Y}_{l,s}^{J^{\prime},m_{J}^{\prime}\dagger}(
\widehat{\bm{r}})
\end{split}
\end{equation}
where we have defined
\begin{equation}
\mathcal{Y}_{l,s}^{J^{\prime},m_{J}^{\prime}}(\widehat{\bm{r}}) = \sum_{m_{l}^{\prime}, m_{s}^{\prime}} C_{l, s, J}^{m_{l}^{\prime}, m_{s}^{\prime}, m_{J}^{\prime}} Y_{l}^{m_{l}^{\prime}}(\widehat{\bm{r}}) \xi_{s}^{m_{s}^{\prime}}.
\end{equation}
Then as the interaction Hamiltonian only connects the $Q\overline{Q}$
component $\psi_{Q\overline{Q}}(\bm{r})$ of the bound
state $A$ and the free $BC$ state, both with the same $J$, $m_{J}$ quantum numbers, we get
\begin{multline}
\bra{\bm{p}_{BC};s_{B},s_{C}
,s,m_{s}} H_{I}\ket{A}\\
=\sum_{l,m_{l}} C_{l, s, J}^{m_{l}, m_{s}, m_{J}} Y_{l}^{m_{l}
}(\widehat{\bm{p}}_{BC})\mathcal{I}_{l}(
p_{BC})
\end{multline}
where we have defined
\begin{equation}
\mathcal{I}_{l}(p_{BC})\equiv\sqrt{\frac
{2}{\pi}}i^{-l}\int \mathrm{d}r \,r^{2}j_{l}(p_{BC
}r)V_{\textup{mix}}^{(BC)}(r)u_{Q\overline{Q}}(
r)
\label{ILP}
\end{equation}
being $u_{Q\overline{Q}}$ the sum of the $Q\overline{Q}$ radial wave functions for the several
$(l_{Q\overline{Q}},s_{Q\overline{Q}})$ channels in the bound state $A$.


Hence we obtain
\begin{multline}
\bra{\bm{p}_{BC};s_{B},m_{s_{B}}
,s_{C},m_{s_{C}}} H_{I}\ket{A}\\
=\sum_{\substack{s,m_{s}, l,m_{l}}}
C_{s_{B}, s_{C}, s}^{m_{s_{B}}, m_{s_{C}}, m_{s}}
C_{l, s, J}^{m_{l}, m_{s}, m_{J}}
Y_{l}^{m_{l}}(\widehat{\bm{p}}_{BC})\mathcal{I}_{l}(p_{BC}) .
\end{multline}
By substituting this in \eqref{rawmcorr},
integrating in spherical coordinates, and using angular momentum algebra it
is easy to show that
\begin{equation}
\mathrm{M}-\mathrm{M}_{A}=\mathcal{P}\!\!\!\int \mathrm{d}p_{BC}\,
\frac{p_{BC}^{2}}{\mathrm{M}-E_{BC}}
\sum_{l,s}\abs{\mathcal{I}_{l}(p_{BC})}^{2} .
\end{equation}
We can go further integrating over $E_{BC}$ instead of over $p_{BC}$ using
\begin{equation}
\frac{\mathrm{d}E_{BC}}{\mathrm{d}p_{BC}}=\frac{p_{BC}
}{m_{B}}+\frac{p_{BC}}{m_{C}}=\frac{p_{BC}}{\mu_{BC}},
\end{equation}
that gives
\begin{equation}
p_{BC}\mathrm{d}p_{BC}=\mu_{BC}\mathrm{d}E_{BC}.
\end{equation}
Thus we obtain
\begin{equation}
\mathrm{M}-\mathrm{M}_{A}=\mathcal{P}\!\!\!\int \mathrm{d}E_{BC}\,\mu_{BC}
\frac{p_{BC}}{\mathrm{M}-E_{BC}}\sum_{l,s}\abs{\mathcal{I}_{l}(p_{BC})}
^{2}
\end{equation}
where
\begin{equation}
p_{BC}=\sqrt{2\mu_{BC}(E_{BC}
-m_{B}-m_{C})}
\end{equation}
and the sum runs over the possible values $l$ and $s$ of $BC$ that couple to
$J^{PC}$.
On the other hand the expression of the width reads
\begin{equation}
\frac{\Gamma}{2}=\pi \mu_{BC} p_{BC}\sum_{l,s}\abs{\mathcal{I}_{l}(p)_{BC}}^{2}\biggr\rvert_{E_{BC}=\mathrm{M}}.
\end{equation}


The generalization of these expressions to the case of one bound state $A$ and
$n$ thresholds with masses below $\mathrm{M}_{A}$ can be done straightforwardly.
So
\begin{equation}
\mathrm{M}-\mathrm{M}_{A}  =
\sum_{j, l_{(j)}, s_{(j)}} \mathcal{P}\!\!\!\int \mathrm{d}E_{(j)}\,\mu_{BC}\frac{p_{(j)}}{\mathrm{M}-E_{(j)
}}\abs{
\mathcal{I}_{l_{(j)}}(p_{(j)})}^{2}
\label{masscorr}
\end{equation}
and
\begin{equation}
\frac{\Gamma}{2}=\sum_{j, l_{(j)}, s_{(j)}}\pi p_{(j)}
\abs{\mathcal{I}_{l_{(
j)}}(p_{(j)})}^{2}\biggr\rvert
_{E_{(j)}=\mathrm{M}}.
\label{width}
\end{equation}


It is important to emphasize that the above expressions are beyond a
perturbative treatment of $H_{I}$, see the appendix, although the mass corrections and
widths obtained are quantitavely similar to those from second order
perturbation theory. Indeed, the second order perturbative correction to the
energy reads
\begin{equation}
\delta\mathrm{M}_{\textup{pert}}^{(k)} = \sum_{j, s} \int \mathrm{d}\bm{p} \, \frac{\abs{
\bra{\Psi_{\textup{free}(j)}^{(\bm{p}, s)}} H_{I} \ket{\Psi_{\textup{bound}\,(N-n)}^{(k)}}}^{2}}{\mathrm{M}_{(N-n)}^{(k)}
-E_{(j)}^{(\bm{p})}} .
\end{equation}
Each term in the sum over $j$ can be integrated separating the principal value and an imaginary contribution from the pole.
Then the principal values and the the pole contributions correspond respectively to the RHS of 
\eqref{masscorr} and \eqref{width} with $\mathrm{M}$ substituted by $\mathrm{M}_{A} = \mathrm{M}_{(N-n)}^{(k)}$.
As in practice
the difference induced by this substitution is quite small
the second order perturbative results yield a quite good approximation. However, as shown in the appendix,
the perturbative correction at fourth order diverges, what invalidates the perturbative treatment.

\section{\label{sec4}Results}

It is important to emphasize that our calculation of mass corrections and
widths is based on the very same form of the interaction Hamiltonian we used to get the bound states.
No new parameters have been introduced. Hence, the comparison to data may
allow us to know the level of accuracy of our description and to think about
possible improvements.

\subsection{Masses}

The calculated mass shifts and corrected masses for the $J^{PC}=1^{--}$ and
$(0,1,2)^{++}$ states below $4.1$~GeV (except the third $1^{--}$ excited state corresponding
to $\psi(4040)$) are presented in
Table~\ref{mass}.

\begin{table}
\begin{ruledtabular}
\begin{tabular}{ccccc}
$J^{PC}$ & $
\mathrm{M}-\mathrm{M}_{A}$ (MeV) & $
\mathrm{M}$ (MeV) & $
\mathrm{M}^{\textup{Expt}}$ (MeV) & Meson\\
\hline
$0^{++}$ 	& 0 & $3508.8$ & $3414.71\pm0.3$ & $\chi_{c0}(1P)$\\
		& $2.0$ & $3920.9$ & $3862$ $_{-32-13}^{+26+40}$ & $\chi_{c0}(3860)$\\
$1^{++}$ 	& $0$ & $3509.8$ & $3510.67\pm0.05$ & $\chi_{c1}(1P)$\\
		& $0$ & $3871.5$ & $3871.69\pm0.17$ & $\chi_{c1}(3872)$\\
$2^{++}$ 	& $0$ & $3508.7$ & $3556.17\pm0.07$ & $\chi_{c2}(1P)$\\
		& $-27.9$ & $3881.1$ & $3922.2\pm1.0$ & $\chi_{c2}(3930)$\\
		& $-2.7$ & $4003.9$ &  & \\
$1^{--}$ 	& $0$ & $3082.4$ & $3096.900\pm0.006$ & $J/\psi(1S)$\\
		& $0$ & $3658.8$ & $3686.097\pm0.010$ & $\psi(2S)$\\
		& $-14.1$ & $3771.7$ & $3778.1\pm0.7$ & $\psi(3770)$
\end{tabular}
\end{ruledtabular}
\caption{\label{mass}Calculated mass corrections, $\mathrm{M} - \mathrm{M}_{A}$, and total masses, M, for
$J^{PC}=1^{--}$ and $(0,1,2)^{++}$ states below $4.1$ GeV. Measured masses $\mathrm{M}^{\textup{Expt}}$ from \cite{PDG20}, corresponding to
the meson assignment in the last column, are also given for comparison.}
\end{table}

Notice that we have included for completeness the first $1^{++}$
excited state that we use for fixing the parameters of the mixing potential.
We have not listed data from $X(3915)$ since its assignment to $0^{++}$ or
$2^{++}$ is not well established. We shall come back to this state later on.

A look at Table~\ref{mass} shows that further mass corrections have to be
implemented for an accurate description of data. In this regard, we have first
checked that the substitution of the nonrelativistic expression of $p_{(j)}$
by the relativistic one does not make any difference. Next we
analyze whether corrections to the Cornell potential could help to
solve, at least in part, the observed discrepancies.

To evaluate these corrections we recall that spin splittings from a single
channel Cornell potential model based on \eqref{CPOT} and \eqref{params} have been calculated \cite{Fei81}.
Hence, for states with approximately a
$100\%$ $c\overline{c}$ probability we can just translate the results from
\cite{Fei81}: $32.4$~MeV for the $1^{--}$ ground state, $-94.1$~MeV for the
$0^{++}$ ground state, $-29.4$~MeV for the $1^{++}$ ground state, and $36.5$~MeV for the $2^{++}$
ground state. This gives total masses for all these ground states differing at most
$30$~MeV from the experimental ones.

For the excited states, with significant meson-meson components, we cannot
perform the same kind of analysis. Instead, spin dependent terms from the
Cornell potential as well as spin dependent terms of the same order from the
mixing and meson-meson potentials should be incorporated to the Schr\"{o}dinger equation. As
this is not an affordable task at present (we do not have any information on
the spin dependence of the mixing potential) we just point out that the
reasonable values of the masses we get for the excited states, as compared to
the ones from experimental candidates, and the values of the spin splittings
from the single channel Cornell potential model \cite{Fei81,Eic04} suggest
that the spin corrections to the masses, apart from the ones related to the
use of the experimental meson masses, should be at most of a few tens of MeV.
This makes us confident of our prediction of a second excited $2^{++}$ state
with a mass about $4$~GeV.

\subsection{Widths}

The calculated widths for $J^{PC}=1^{--}$ and $(0,1,2)^{++}$
states below $4.1$ GeV and above the $D\overline{D}$ threshold (except the third $1^{--}$ excited state corresponding
to $\psi(4040)$) are compiled
and compared to existing data in Table~\ref{Tablewidths}.

\begin{table*}
\begin{ruledtabular}
\begin{tabular}{ccccccccc}
$J^{PC}$ & $\mathrm{M}$ (MeV) &
$\Gamma_{D\overline{D}}$ (MeV)
& $\Gamma_{D\overline{D}^{\ast}}$ (MeV)
& $\Gamma_{D_{s}\overline{D}_{s}}$ (MeV)
& $\Gamma_{D^{\ast}\overline{D}^{\ast}}$ (MeV)
& $\Gamma_{\textup{total}}^{\textup{Theor}}$ (MeV)
& $\Gamma_{\textup{total}}^{\textup{Expt}}$ (MeV)
& Meson\\
\hline
$0^{++}$ 	& $3920.9$ & $0.6$ &  &  &  & $0.6$ & $201$ $_{-67-82}^{+154+88}$ & $\chi_{c0}(3860)$\\
$1^{++}$ 	& $3871.7$ &  &  &  &  & 0  & $<1.2$ & $\chi_{c1}(3872)$\\
$2^{++}$ 	& $3881.1$ & $49.5$ & $0.4$ &  &  & $49.9$ & $35.3\pm2.8$ & $\chi_{c2}(3930)$\\
		& $4003.9$ & $4.8$ & $6.3$ & $3.5$ &  & $14.5$ &  & \\
$1^{--}$	& $3771.7$ & $20.2$ & & & & $20.2$ & $27.2\pm1.0$ & $\psi(3770)$ \\
\end{tabular}
\end{ruledtabular}
\caption{\label{Tablewidths}Calculated widths to open-charm meson-meson channels.
Experimental total widths $\Gamma_\textup{total}^{\textup{Expt}}$ from
\cite{PDG20}, corresponding to the meson assignment in the last column, are
quoted for comparison.}
\end{table*}
A look at it shows that the results for the total widths are fairly
good. So, for $\psi(3770)$ the calculated $\Gamma
_{D\overline{D}}$ is about a $10\%$ below the experimental interval (we have
used that $\frac{\Gamma_{D\overline{D}}^{\textup{Expt}}}{\Gamma_{\textup{total}}^{\textup{Expt}}
}=0.93_{-9}^{+8}$ for a better comparison$)$, and for $\chi_{c2}(
3930)$ about a $30\%$ above data (for
$\chi_{c0}(3860)$ better data are needed for a sensible
comparison). Moreover, we predict for $\chi_{c2}(3930)$ a
dominant decay to $D\overline{D}$, as experimentally observed. We can go
further since we know for $\chi_{c2}(3930)$ the experimental
ratio
\begin{equation}
\frac{\Gamma_{D\overline{D}}^{\textup{Expt}} \Gamma
_{\gamma\gamma}^{\textup{Expt}}}{\Gamma_{\textup{total}}^{\textup{Expt}}}
=0.21\pm0.04~\text{keV}.
\label{measuredratio}
\end{equation}
We can calculate this ratio from $\Gamma_{D\overline{D}}$ and $\Gamma_{\textup{total}}^{\textup{Theor}}$ in Table~\ref{Tablewidths}, and
$\Gamma(\chi_{c2}(3930)\rightarrow\gamma\gamma)$, which can be estimated from \cite{Kwo88} as
\begin{equation}
\frac{\Gamma(\chi_{c2}(3930)\rightarrow\gamma
\gamma)}{\Gamma(\chi_{c2}(1P)\rightarrow
\gamma\gamma)}\approx\frac{\abs{(u_{c\overline{c}
}^{\prime}(0))_{\chi_{c2}(3930)
}}^{2}}{\abs{(u_{c\overline{c}}^{\prime}(
0))_{\chi_{c2}(1P)}}^{2}}= 0.80
\end{equation}
where $u_{c\overline{c}}^{\prime}(0)$ stands for the derivative
of the $c\overline{c}$ radial wave function at the origin.

Then, using the measured value
\begin{equation}
\Gamma(\chi_{c2}(1P)\rightarrow\gamma
\gamma)^{\textup{Expt}}= 0.57\pm 0.05~\text{keV}
\end{equation}
we get
\begin{equation}
\frac{\Gamma_{D\overline{D}}^{\textup{Theor}} \Gamma
_{\gamma\gamma}^{\textup{Theor}}}{\Gamma_{\textup{total}}^{\textup{Theor}}}
\approx 0.45 \pm 0.04~\text{keV}
\end{equation}
which is a factor 2 bigger than data.

These results for $\chi_{c2}(3930)$ seem to indicate that we
overestimate the probability of the $c\overline{c}$ component since a (by
hand) modest decrease in it makes the total width as well as the calculated
ratio to be close to data. This overestimation may have to do with the
previously mentioned need of refining our approach for a correct description
of $\psi(4040)$.

This refinement should also affect our predicted width for the second excited
$2^{++}$ state and first excited $0^{++}$
state. Anyhow, we think we can safely predict in both cases widths of at most
a few tens of MeV.

It is also worth to comment about the $X(3915)$. In Ref.~\cite{PDG20} $D^{0}\overline{D}^{\ast0}$ appears as a
possible decay mode. If confirmed $X(3915)$ would be a $2^{++}$
resonance since this decay mode is forbidden for $0^{++}$. The $2^{++}$
assignment would mean its identification with $\chi_{c2}(3930)
$. This is perfectly compatible with the average measured masses,
$3918.4\pm1.9$~MeV for $X(3915)$ and $3922.2\pm1.0$~MeV for
$\chi_{c2}(3930)$. Moreover, according to our theoretical
description of $\chi_{c2}(3930)$ the discovery channel for
$X(3915)$, $J/\psi \, \omega$, would be a natural decay mode of
$\chi_{c2}(3930)$ from its $D^{\ast}\overline{D}^{\ast}$
component via quark exchange. Hence, the confirmation of the experimental
observation of the $D\overline{D}^{\ast}$ decay mode for $X(
3915)$ could definitely settle the question about its $J^{PC}$ quantum
numbers and its possible identification with $\chi_{c2}(3930)$.

To conclude this section, let us point out that the effects of open-charm meson-meson
on charmonium properties has been extensively investigated in the
literature, see for example \cite{Eic80,Eic04,Eic06,Bar05,Bar08,Pen07,Dan10,Fer13,Ort18,vanB21}.
Although our diabatic results for mass corrections and widths cannot be
directly compared to the ones calculated within different frameworks, some
comments to contrast our approach with previous ones are in order.

A major difference of our diabatic approach with respecto to all these studies
is the incorporation of a lattice-based form of the mixing potential. The
diabatic formalism allows to connect the mixing potential with unquenched
lattice QCD results of string breaking, that is the actual physical mechanism
underlying the $Q\overline{Q}$--meson-meson mixing. Conversely, in \cite{Eic80,Eic04,Eic06}
the mixing potential was derived from the Cornell
potential whereas in most of the other studies it was generated from a
phenomenological $^{3}\!P_{0}$ model, in both cases with no clear connection
to QCD. In this respect, it is important to emphasize that our parametrization
of the mixing potential, which is instrumental for the calculation of masses
and widths, can be straightforwardly incorporated to these studies.

Another important difference with some previous studies evaluating mass
corrections and widths from open-charm meson-meson loops is the complete
nonperturbative character of our calculation. As a matter of fact we have shown
that a perturbative treatment to the lowest order is untenable for it is
divergent to the next nonvanishing order. 

Its complete nonperturbative character and the incorporation of the (lattice)
QCD mixing potential are the very distinctive signatures of our diabatic
approach. Certainly, at its present stage, it needs refinements for an
accurate description of data, but these signatures have to be present in any
description of quarkoniumlike systems from QCD.

\section{\label{sec5}Summary}

In this article, the second of a series on the diabatic approach in QCD, we
have completed the study of $(0,1,2)^{++}$ and $1^{--}$ charmoniumlike resonances
below $4.1$~GeV (except $\psi(4040)$). In the
first article of the series we developed a bound state approximation to a
charmoniumlike resonance involving the coupling of $c\overline{c}$ and
meson-meson channels with thresholds above the
mass of the resonance. The (diabatic) potential matrix in the resulting
Schr\"{o}dinger equation for the $c\overline{c}$ and meson-meson components
was obtained from an educated guess of the corresponding static energies to be calculated in lattice QCD.
This is a major difference with previous quark model studies where the non
diagonal matrix elements mixing the $c\overline{c}$ and meson-meson
configurations did not have a clear connection to QCD.

In this second article, we have proceeded to a nonperturbative calculation of
mass corrections and widths from meson-meson thresholds below the mass of the resonance. It is worth to remark that there is not
any new free parameter in this calculation. Hence, the predicted results are
directly testing our description. In this regard, we have shown that the
additional consideration of spin dependent terms in the potential matrix could
give account of the discrepancies between the calculated masses and data. 

As for the calculated total widths, they are in fair agreement with data. A
more detailed comparison with the scarce available data involving partial widths
shows that the discrepancies with data could be due to the need of refining our
approach beyond our corrected bound state treatment for a better wave function description.

The reasonable results we get for masses and widths make us consider as quite
robust the prediction of the existence of a second excited $2^{++}$
state, being dominantly a $c\overline{c}$ $f$-wave, with a mass about $4$~GeV,
and width of a few tens of~MeV. The experimental discovery
of this resonance would provide strong support to our description.

Therefore, from our analysis of $1^{--}$ and $(0,1,2)^{++}$
charmoniumlike states below $4.1$ GeV, we may tentatively conclude that the
diabatic approach in QCD is an adequate framework for a complete treatment of
conventional and unconventional charmoniumlike resonances.

\begin{acknowledgments}
This work has been supported by MINECO of Spain and EU Feder Grant No.\ FPA2016-77177-C2-1-P,
by EU Horizon 2020 Grant No.\ 824093 (STRONG-2020) and by PID2019-105439GB-C21. R.~B. acknowledges a FPI
fellowship from MICIU of Spain under Grant No.~BES-2017-079860.
\end{acknowledgments}

\appendix*

\section{Perturbative corrections}

Starting from the unperturbed Hamiltonian $H_{0}$ whose eigenvectors
$\ket{\Psi^{(0)}}$ and eigenvalues $E^{(0)}$ are known,
\begin{equation}
H_{0}\ket{\Psi^{(0)}}=E^{(0)}\ket{\Psi^{(0)}},
\end{equation}
the perturbation equation of order $i$ reads
\begin{equation}
H_{0}\ket{\Psi^{(i)}}+H_{I}\ket{\Psi^{(i-1)}}=\sum_{j=0}^{i}E^{(j)}%
\ket{\Psi^{(i - j)}},\label{perteq}%
\end{equation}
where $H_{I}$ is the interaction Hamiltonian treated as a perturbation and
$\ket{\Psi^{(i)}}$, $E^{(i)}$ are the the $i$-th order perturbative
corrections to the wave function and energy respectively.

Let us now focus, for the sake of simplicity, on the perturbative solution of
the diabatic problem with only one meson-meson threshold. In this case the
unperturbed solutions are the discrete bound states, and the the continuum of
meson-meson states corresponding to the threshold:
\begin{align}
H_{0}\ket{\Psi_{k}^{(0)}} &  =M_{k}^{(0)}\ket{\Psi_{k}^{(0)}}\\
H_{0}\ket{\Psi_{\bm{p}}^{(0)}} &  =E_{\bm{p}}^{(0)}\ket{\Psi_{\bm{p}}^{(0)}}
\end{align}
where $M_{k}^{(0)}$ is the mass of the $k$-th bound state and $E_{\bm{p}}%
^{(0)}$ is the c.o.m.\ meson-meson energy, given as usual by
\begin{equation}
E_{\bm{p}}^{(0)}=\frac{p^{2}}{2\mu}+T
\end{equation}
where $\bm{p}$ is the relative momentum, $p$ its modulus, $\mu$ the reduced
mass, and $T$ the threshold mass.

We now set off to calculate the perturbative corrections
$\ket{\Psi_{k}^{(i)}}$ and $M_{k}^{(i)}$ of some discrete state labeled by
$k$, assuming for simplicity that the unperturbed discrete spectrum is nondegenerate.

To clearly distinguish between the unperturbed bound states and unperturbed
free meson-meson states, let us introduce the notation
\begin{equation}
\ket{\psi_{k}}\equiv\ket{\Psi_{k}^{(0)}}\qquad\ket{\varphi_{\bm{p}}}\equiv
\ket{\Psi_{\bm{p}}^{(0)}}.
\end{equation}
Let us note that in perturbation theory, for $i>0$ one has
$\braket{\psi_{k} \vert \Psi_{k}^{(i)}}=0$ by construction.

By multiplying Eq.~\eqref{perteq} for $i=1$ on the left by $\bra{\psi_{k}}$,
and taking into account that the interaction hamiltonian only connects bound
states and continuum states, it can be easily shown that $M_{k}^{(1)}=0$,
i.e., the first order correction to the mass vanishes. On the other
hand, by multiplying Eq.~\eqref{perteq} for $i=1$ on the left by
$\bra{\psi_{k^{\prime}}}$ with $k^{\prime}\neq k$, it is straightforward to
show that $\braket{\psi_{k^{\prime}} \vert \Psi_{k}^{(1)}} = 0$. This is, there is no first order
correction to the wave function from other discrete states. Actually the only
first order correction to the wave function comes form meson-meson states, and
can be calculated by multiplying Eq.~\eqref{perteq} for $i=1$ on the left by
$\bra{\varphi_{\bm{p}}}$:
\begin{equation}
E_{\bm{p}}^{(0)}%
\braket{\varphi_{\bm{p}} \vert \Psi_{k}^{(1)}}+\bra{\varphi_{\bm{p}}}H_{I}%
\ket{\psi_{k}}=M_{k}^{(0)}%
\braket{\varphi_{\bm{p}} \vert \Psi_{k}^{(1)}}.\label{probleq1}%
\end{equation}
so that we can express
\begin{equation}
\braket{\varphi_{\bm{p}} \vert \Psi_{k}^{(1)}}=\frac
{\bra{\varphi_{\bm{p}}}H_{I}\ket{\psi_{k}}}{M_{k}^{(0)}-E_{\bm{p}}%
^{(0)}+i\epsilon},
\end{equation}
where as customary we have introduced an imaginary factor $i\epsilon$ to deal
with the case $E_{\bm{p}}^{(0)}=M_{k}^{(0)}$. 

Higher order corrections are calculated exactly in the same way as the first
order ones. It is sufficient to multiply the perturbation equation
\eqref{perteq} of order $i$ by $\bra{\psi_{k}}$, $\bra{\psi_{k^{\prime}}}$
with $k^{\prime} \ne k$, and $\bra{\varphi_{\bm{p}}}$ to obtain the $i$-th
order mass correction and wave function corrections respectively. Thus the
second order mass correction reads
\begin{equation}
M_{k}^{(2)} = \bra{\psi_{k}} H_{I} \ket{\Psi_{k}^{(1)}} = \int\mathrm{d}\bm{p}
\frac{\abs{\bra{\varphi_{\bm{p}}} H_{I} \ket{\psi_{k}}}^{2}}{M_{k}^{(0)} -
E_{\bm{p}}^{(0)} + i\epsilon}.
\end{equation}
This term can be physically interpreted as a one meson-meson loop correction
to the mass. Note that as $\epsilon$ goes to zero the simple pole of the
integrand moves on to the real axis, at $E_{\bm{p}}^{(0)} = M_{k}^{(0)}$. For
states above threshold we have $M_{k}^{(0)} > T$, what causes the pole to fall
within the integration range. Then for $\epsilon\to0$ the integral splits in a
real principal part corresponding to the mass shift and an imaginary
contribution that is proportional to the decay width of the quarkonium state
to a meson-meson pair. More precisely, it can be shown that the imaginary
contribution from this one loop correction is exactly half the decay amplitude
calculated at tree level, in accordance with the optical theorem.

The second order correction to the wave function is calculated from
\begin{equation}
M_{k^{\prime}}^{(0)} \braket{\psi_{k^{\prime}} \vert \Psi_{k}^{(2)}} +
\bra{\psi_{k^{\prime}}} H_{I} \ket{\Psi_{k}^{(1)}} = M_{k}^{(0)}
\braket{\psi_{k^{\prime}} \vert \Psi_{k}^{(2)}},
\end{equation}
yielding
\begin{equation}%
\begin{split}
\braket{\psi_{k^{\prime}} \vert \Psi_{k}^{(2)}}  & = \frac
{\bra{\psi_{k^{\prime}}} H_{I} \ket{\Psi_{k}^{(1)}}}{M_{k}^{(0)} -
M_{k^{\prime}}^{(0)}}\\
& = \int\mathrm{d}\bm{p} \frac{\bra{\psi_{k^{\prime}}} H_{I}
\ket{\varphi_{\bm{p}}}}{M_{k}^{(0)} - M_{k^{\prime}}^{(0)}} \frac
{\bra{\varphi_{\bm{p}}} H_{I} \ket{\psi_{k}}}{M_{k}^{(0)} - E_{\bm{p}}^{(0)} +
i \epsilon},
\end{split}
\end{equation}
what can be physically interpreted as a mixing between quarkonium states
mediated by one meson-meson loop.

At third order the mass correction vanishes, $M_{k}^{(3)}=0$, as
can be easily shown. The only nonvanishing contribution, the wave function
correction from the continuum states, is derived from
\begin{multline}
E_{\bm{p}}^{(0)} \braket{\varphi_{\bm{p}} \vert \Psi_{k}^{(3)}} +
\bra{\varphi_{\bm{p}}} H_{I} \ket{\Psi_{k}^{(2)}}\\
= M_{k}^{(0)} \braket{\varphi_{\bm{p}} \vert \Psi_{k}^{(3)}} + M_{k}^{(2)}
\braket{\varphi_{\bm{p}} \vert \Psi_{k}^{(1)}}
\end{multline}
what gives
\begin{equation}
\braket{\varphi_{\bm{p}} \vert \Psi_{k}^{(3)}} = \frac{\bra{\varphi_{\bm{p}}}
H_{I} \ket{\Psi_{k}^{(2)}}}{M_{k}^{(0)} - E_{\bm{p}}^{(0)} + i \epsilon} -
M_{k}^{(2)} \frac{\braket{\varphi_{\bm{p}} \vert \Psi_{k}^{(1)}}}{M_{k}^{(0)}
- E_{\bm{p}}^{(0)} + i \epsilon}.
\end{equation}

Finally, the fourth order mass correction reads
\begin{nobotrulewidetext}
\begin{multline}
M_{k}^{(4)} = \bra{\psi_{k}} H_{I} \ket{\Psi_{k}^{(3)}} = \sum_{k^{\prime} \ne
k} \iint\mathrm{d} \bm{p} \, \mathrm{d} \bm{p}^{\prime} \frac{\bra{\psi_{k}}
H_{I}\ket{\varphi_{\bm{p}}} \bra{\varphi_{\bm{p}}} H_{I}
\ket{\psi_{k^{\prime}}} \bra{\psi_{k^{\prime}}} H_{I}
\ket{\varphi_{\bm{p}^{\prime}}} \bra{\varphi_{\bm{p}^{\prime}}} H_{I}
\ket{\psi_{k}}}{(M_{k}^{(0)} - E_{\bm{p}}^{(0)} + i \epsilon) (M_{k}^{(0)} -
M_{k^{\prime}}^{(0)}) (M_{k}^{(0)} - E_{\bm{p}^{\prime}}^{(0)} + i \epsilon
)}\\
- \left(  \int\mathrm{d}\bm{p}^{\prime} \frac
{\abs{\bra{\varphi_{\bm{p^{\prime}}}} H_{I} \ket{\psi_{k}}}^{2}}{M_{k}^{(0)} -
E_{\bm{p}^{\prime}}^{(0)} + i \epsilon} \right)  \cdot\left(  \int
\mathrm{d}\bm{p} \frac{\abs{\bra{\varphi_{\bm{p}}} H_{I} \ket{\psi_{k}}}^{2}%
}{(M_{k}^{(0)} - E_{\bm{p}}^{(0)} + i \epsilon)^{2}} \right)
\end{multline}
where it can be seen that when $M_{k}^{(0)} > T$ the last integral of the
second line is divergent as $\epsilon\to0$, since the integrand has a double pole.

The divergence of the fourth order mass correction reflects the fact that the
presence of continuum states with energy equal to the mass of some discrete
states invalidates the perturbative treatment for these states.
\end{nobotrulewidetext}

\bibliography{masswidbib}

\end{document}